# On Orthogonality of Latin Squares

R.N.Mohan[1], Moon Ho Lee[2], and Subash Shree Pokhrel[3]
Sir CRR Institute of Mathematics, Sir CRR College Campus, Eluru-534007, AP.,India[1]
Institute of Information &Communication, Chonbuk National University, South Korea[2,3]
Email: mohan420914@yahoo.com[1], moonho@chonbuk.ac.kr[2], subash21us@yahoo.com[3]

**Abstract:** A Latin square arrangement is an arrangement of s symbols in s rows and s columns, such that every symbol occurs once in each row and each column. When two Latin squares of same order superimposed on one another, then in the resultant array every ordered pair of symbols occurs exactly once, then the two Latin squares are said to be orthogonal. A frequency square M of type F (n; λ) is an n x n, matrix over an m-set S, where n=mλ, such that every element of S occurs exactly λ times in each row and each column of M. Two frequency squares of the same type over S are orthogonal if one is superimposed on the other each element of S x S appears $\lambda^2$ times. These two concepts lead to a third concept that is if t-orthogonal Latin squares of order n, from a set S of Latin squares, are superimposed, then in the resultant array, each *t*-tuple occurs exactly once. If at all it is possible then how to construct them and that is the genesis of the paper. In this paper while generating Latin square (Latin rectangles), a new concept called t-orthogonality over the set of Latin structure has been discussed and their constructions have been given.

**1. Introduction** From Latin square enumeration for example, refer to Wilson [27], we can know that how many Latin squares can exist for a given s the order of the Latin square, but question is of their construction. There seems to be no good algorithm for constructing a random Latin square. One natural approach to counting and constructing Latin squares is to do it one row at a time, there by defining, "Latin rectangles", and then try to obtain exact and asymptotic formulae, using the structural properties of the under lying templates. (for other references and results see Denes and Keedwell [5, 6]). Latin Squares were invented and studied by Euler [7] in 1782.

A Latin square arrangement is an arrangement of s symbol in $s^2$ cells arranged in s rows and s columns such that each symbol occurs once in each row and in each column. This s is called the order of the Latin square. Two Latin squares of the same order s when superimposed on one another and if each pair of symbols in the resultant array occurs only once they are called orthogonal. On a given set of N Latin squares any two Latin squares are Orthogonal then the set is called mutually orthogonal latin squares (MOLS) of order s. The cardinality of this set N is denoted by $O(N) \leq s-1$.

A frequency square M of type F(n; λ) is an n x n, matrix over an m-set S, where n = mλ, such that every element of S occurs exactly λ times in each row and each column of M. Two frequency squares of the same type over S are orthogonal if one is superimposed on the other each element of S x S appears $\lambda^2$ times. These two concepts lead to a third concept that is if t-orthogonal Latin squares of order n, from a set S of Latin squares, are superimposed, then in the resultant array, each *t*-tuple occurs exactly once. If it is possible then how to construct them is the theme of our present paper. In this paper we gave some methods of construction of those *t*-orthogonal Latin squares.

---------------------------------------

Much of the work on Latin squares has been done by various authors for example Bose [1], Bose, Shrikhande and Parker [2], Tarry [23], Wallis [25, 26], Mann [15] and so on, who gave the methods of construction of these MOLS in various ways.

In the present paper, while giving various methods of generating Latin squares (rectangles), we purpose a new concept called t-orthogonality over a set of these Latin squares and rectangles of the same order. Also we will make some discussions with the comparative illustrations of 2-orthogonality (classical) to t-orthogonality and end with a tabulation of the Latin squares of order s = 1, 2,…, 45 and their orthogonalities and methods of construction, for ready reference to researchers.

## 2. Main result:

**Definition 2.1.** If *t* Latin squares of the same order s, $2 \leq t \leq s$ from a set of Latin squares are superimposed one on another and in a cell, the ordered *t*-tuple occurs once and only once in the resultant array, then they are *t*-orthogonal and the set is called the set of mutually *t*- orthogonal Latin squares.
Let $t(N)$ denotes the maximum number *t*-orthogonal Latin squares each of order s. Now the classical orthogonality can be renamed as 2-orthogonality or as orthogonality only. So the notation for this may be retained as $O(N)$) itself.

**Conjecture 2.1:** Every set of mutually *t*-orthogonal Latin squares is also (*t*+1)-orthogonal, but the converse is not true, where $2 \leq t \leq t+1 \leq s$, and s is the order of the Latin squares.

**An explanation:** Let $(\theta_1, \theta_2, \theta_3,...,\theta_t, \alpha)$, $(\theta_1, \theta_2, \theta_3,...,\theta_t, \beta)$ be two distinct *t*-tuples. On addition of $\alpha$ and $\beta$ the orthogonality establishes and the deletion of $\alpha$ and $\beta$ the orthogonality fails. Two distinct 2-tuples will never become equal by the addition of any number of same elements to them. And the orthogonality maintains. If some *t*-tuple repeats, then by addition of another elements, if they are the same, the (*t*+1)-tuple also repeats, if they are distinct, then those (*t*+1) tuples will be distinct. By such additions, the *t*-tuple in repetition, at certain stage may cease to repeat and thus paves the way for t-orthogonality.

**Note 2.1:** In Macneish [14, p.224], it was stated that, "if the square of index n, 2 does not exist, then the square of index n, k, for k>2 can't exist". This is not true. See Note that is given below.
This concept of generalized orthogonality is not to be confused with the Macneish [14], concept on Latin square of index k. The two concepts are basically different. Macneish [14], speaks about Latin square with certain symbols (may be t-tuples) as entries but here, we speak of t-orthogonality between t given Latin squares from a set of Latin squares.

**Note 2.2:** The conjecture in the note above may mislead the reader challenging the very utility of these results. Even though we have some Latin squares, which are not *t*-orthogonal, we can establish that they are (*t*+1)-orthogonal. For example for Latin square of order 6, there exists no orthogonal mate, but *t*-orthogonal mate exists. For Latin squares of order 8, the number of MOLS is 5. But we have *t* (8) =8. Besides, this concept is having an application in certain network systems (can be seen in the same issue elsewhere).

**Note 2.3:** Many of the constructions of mutually *t*-orthogonal Latin squares are yet to be known.

**Note 2.4:** In the *t*-tuple, either all the t elements must be distinct or must be the same, but every *t*-tuple must be unique in the resultant array.

**Proposition 2.1.** When s is prime or (s+1) is prime then there exists a set of mutually *t*-orthogonal Latin squares of order s.

In Wallis [25], the following result was given.

**Theorem 2.1:** If there are k mutually 2-orthogonal Latin squares of order s, $s \geq 3$ then k<s.

Analogously, we can have the following result.

**Proposition 2.2.** If there are k mutually *t*-orthogonal Latin squares of order s, $s \geq 3$ then k = s-1, when s is prime and k = s when (s+1) is prime. Otherwise we can state this as follows.

**Proposition 2.3.** If there exists a set N of mutually t-orthogonal Latin squares each of order s then
$$t(N) \leq s$$
When (s+1) is prime, then $t(N)$ attains its upper bound.

## 3. Methods of construction:

Recently Mohan [17] identified a family of matrices and established its usefulness in the construction of designs, graphs and codes.

In fact by adopting the formula, which was originally used in the construction of mutually orthogonal Latin squares and after further generalizing it, that is being made use of for the above purpose. The formula is
$$M_n = (a_{ij}^h)$$
where
$$a_{ij}^h = (d_i \oplus d_h d_j) \bmod n \qquad (1)$$
For various values of $d_i, d_h$ and $d_j$ and n by defining $\oplus$ in different ways, for example by taking ordinary (+), subtraction (-) and multiplication many matrices have been constructed (cf. [17]). Now by taking formula and by construction some of these possibilities in (1) we have

**Case (A):** When $d_i = i$, $d_j = j$, $d_h = h$, and $\oplus$ is (+) we get $(i + hj) \bmod n$. When (h, n) = 1, we get a Latin square.

**Case (B):** When $\oplus$ is taken as (-), we can have $(i - hj) \bmod n$ when (h, n) = 1. Then also we get a Latin square. But these tow types Case (A) and Case (B) are combinatorially equivalent. Hence we consider Case (A) only.

**Case (C):** When $d_i = i$, $d_j = j$, $d_h = h$ and $\oplus$ is (.) we have $(i.hj) \bmod n$. Again by suitably defining $d_i, d_h, d_j$, we can also get
$$(i-1)(j-1) \bmod n, [1+(i-1)(j-1)] \bmod n,$$

which are all combinatorially equivalent. So we consider $(i.j) \mod n$ only. And now we establish that all these are useful for the construction of the sets of Latin squares (rectangles) with the property of *t*-orthogonality.

**Note 3.1:** In these cases, if we take i = j, we get Latin squares, and if we take i<j or j<i we get Latin rectangles.

**Proposition 3.1.** If (s+1) is a prime and if $1 < i \leq s$ and $1 < j \leq s$, then $(i.j) \mod (s+1)$ given out a Latin square and when $(i \neq j$, a Latin rectangle.

**Proof:** Suppose for two distinct columns $j_1, j_2$, in the same row we have the elements as

$$a_{ij_1}, a_{ij_2}$$

and now suppose

$$a_{ij_1} = a_{ij_2} \therefore (i.j_1) = (i.j_2) \therefore j_1 = j_2$$

It is a contradiction. There fore $a_{ij_1} \neq a_{ij_2}$ again suppose for distinct rows $i_1, i_2$ in the same column we have the element as

$$a_{i_1,j}, a_{i_2,j}$$

And now if

$$a_{i_1,j} = a_{i_2,j} \therefore (i_1.j) = (i_2.j) \therefore i_1 = i_2$$

Again a contradiction. Therefore

$$a_{i_1,j} \neq a_{i_2,j}$$

Hence it is a Latin square.

**Proposition 3.2.** If s is prime and $1 < i \leq s$, and $1 < j \leq s$. Then $(i + hj) \mod s$, $(h,s) = 1$, gives out a Latin square when $i = j$ and Latin rectangle when $i \neq j$.

**Proof:** Suppose for two distinct columns $j_1$ and $j_2$ in the same row we have the elements as $a_{i,j_1}, a_{i,j_2}$. Let us suppose that

$$a_{i,j_1} = a_{i,j_2} \therefore (i + hj_1) = (i + hj_2) \therefore j_1 = j_2$$

It is a contradiction. Therefore

$$a_{i,j_1} \neq a_{i,j_2}$$

Similarly for two distinct rows $i_1$ and $i_2$ and in the same columns, let the elements be

$$a_{i_1,j} = a_{i_2,j} \therefore (i_1 + hj) = (i_2 + hj) \therefore i_1 = i_2$$

It is a contradiction. Hence $a_{i_1,j} \neq a_{i_2,j}$ so it is a Latin square.

**Note 3.2:** For the other cases namely $(i - hj) \mod s$ when s is prime, and $(h,s) = 1$ $[(i-1)(j-1)] \mod s, [1 + (i-1)(j-1) \mod s]$ when (s+1) is prime, we can prove analogously that they also give Latin squares.

**Note 3.3:** When (s+1) is not a prime, by the following formula $(1+at) \mod (s+1))$, the elements in the columns of $\{(i.j) \mod(s+1)\}$, repeat t-times where $(x_i,(s+1))=t$, where $x_j=(j+1)$ column and where $j=0,1,2,...,s-1$, $x_j=1,2,...,s$ and $a=0,1,2,...,s-1$.

## 4. Discussion:

Here we discuss with the illustrative comparison of 2-orthogonality of $t$-orthogonality in Lain squares of order s = 1, 2, …, 41, one by one.

Primarily there are four types of numbers for our purpose. They are (1) when s is prime (2) when (s+1) is prime (3) when s is a prime power and (4) all the remaining integer values of s, say some composite numbers.

**Type I:** This is of the type when s is prime i.e. when

$$s = 1,2,3,5,7,11,13,17,19,23,29,31,37,41\ldots$$

**Examples .4.1**.

i) For s=1, without loss of generality we can say $t(N) = \infty$. In this case we get $t$-tuple of unities and there exists only one such $t$-tuple. And there are infinite singleton matrices, consisting of unit elements.

ii) For s = 2, it does not have a t-orthogonal mate $t(2) = 1$. Even though the number of Latin squares of order 2 are two, they are not orthogonal. $\begin{bmatrix} 1 & 2 \\ 2 & 1 \end{bmatrix}\begin{bmatrix} 2 & 1 \\ 1 & 2 \end{bmatrix}$ are the two Latin squares of order 2, and on superimposition where $\begin{bmatrix} 12 & 21 \\ 21 & 12 \end{bmatrix}$, each pairs occurs twice.

iii) For s = 3, we cannot construct 3-orthogonal but we can get 2-orthogonal. Here there is possibility of having ordered 2-tuples with the same elements also, and thus having 9 cells and 9 distinct ordered pairs. By using the formula $(a_{ij}^h) = (i + hj)\mod 3, h = 1,2$

$$\begin{array}{cc} h = 1 & h = 2 \\ i = 1,2,3 & i = 1,2,3 \\ j = 1,2,3 & j = 1,2,3 \end{array}$$

$$\begin{array}{cc} A_1 = 231 & A_2 = 321 \\ 312 & 132 \\ 123 & 213 \end{array}$$

$$(A_1 A_2) = \begin{array}{ccc} (2,3) & (3,2) & (1,1) \\ (3,1) & (1,3) & (2,2) \\ (1,2) & (2,1) & (3,3) \end{array}$$

, which is 2-orthogonal.

Hereafter for all the number of s = 5, 7, 11, 13, … we can use the method given in Case (A) and Case (B) i.e.
$$a_{ij}^h = (i \pm hj)\mod s,$$
where $1 \le h \le s-1$ and $(h, s) = 1$, this is always true since s is prime. With this we can construct a set of $t$-orthogonal Latin squares of order s, where as $t(N) = s - 1$.

As said earlier since $(i - hj) \mod s$, gives the similar type of matrices as in $(i + hj) \mod s$, hence we consider only $(i + hj) \mod s$ through out.

iv) for s = 5, by using $(i + hj) \mod 5, h = 1,2,3,4$.

$$h = 1$$
$$i = 1,2,3,4,5$$
$$j = 1,2,3,4,5$$

$$A_1 = \begin{matrix} 2 & 3 & 4 & 5 & 1 \\ 3 & 4 & 5 & 1 & 2 \\ 4 & 5 & 1 & 2 & 3 \\ 5 & 1 & 2 & 3 & 4 \\ 1 & 2 & 3 & 4 & 5 \end{matrix}$$

$$h = 2$$
$$i = 1,2,3,4,5$$
$$j = 2,4,1,3,5$$

$$A_2 = \begin{matrix} 3 & 5 & 2 & 4 & 1 \\ 4 & 1 & 3 & 5 & 2 \\ 5 & 2 & 4 & 1 & 3 \\ 1 & 3 & 5 & 2 & 4 \\ 2 & 4 & 1 & 3 & 5 \end{matrix}$$

$$h = 3$$
$$i = 1,2,3,4,5$$
$$j = 3,1,4,2,5$$

$$A_3 = \begin{matrix} 4 & 2 & 5 & 3 & 1 \\ 5 & 3 & 1 & 4 & 2 \\ 1 & 4 & 2 & 5 & 3 \\ 2 & 5 & 3 & 1 & 4 \\ 3 & 1 & 4 & 2 & 5 \end{matrix}$$

$$h = 4$$
$$i = 1,2,3,4,5$$
$$j = 4,3,2,1,5$$

$$A_4 = \begin{matrix} 5 & 4 & 3 & 2 & 1 \\ 1 & 5 & 4 & 3 & 2 \\ 2 & 1 & 5 & 4 & 3 \\ 3 & 2 & 1 & 5 & 4 \\ 4 & 3 & 2 & 1 & 5 \end{matrix}$$

Since s = 5, a prime number. They form 2, 3, 4-orthogonal

(2345)  (3524)  (4253)  (5432)  (1111)
(3451)  (4135)  (5314)  (1543)  (2222)
(4512)  (5241)  (1425)  (2154)  (3333)
(5123)  (1352)  (2531)  (3215)  (4444)
(1234)  (2413)  (3142)  (4321)  (5555)

v) for s = 7 we use $(i + hj) \mod 7, h = 1,2,3,4,5,6$

| $h = 1$ | $h = 2$ |
|---|---|
| $i = 1,2,3,4,5,6,7$ | $i = 1,2,3,4,5,6,7$ |
| $j = 1,2,3,4,5,6,7$ | $j = 2,4,6,1,3,5,7$ |

$$A_1 = \begin{matrix} 2 & 3 & 4 & 5 & 6 & 7 & 1 \\ 3 & 4 & 5 & 6 & 7 & 1 & 2 \\ 4 & 5 & 6 & 7 & 1 & 2 & 3 \\ 5 & 6 & 7 & 1 & 2 & 3 & 4 \\ 6 & 7 & 1 & 2 & 3 & 4 & 5 \\ 7 & 1 & 2 & 3 & 4 & 5 & 6 \\ 1 & 2 & 3 & 4 & 5 & 6 & 7 \end{matrix} \qquad A_2 = \begin{matrix} 3 & 5 & 7 & 2 & 4 & 6 & 1 \\ 4 & 6 & 1 & 3 & 5 & 7 & 2 \\ 5 & 7 & 2 & 4 & 6 & 1 & 3 \\ 6 & 1 & 3 & 5 & 7 & 2 & 4 \\ 7 & 2 & 4 & 6 & 1 & 3 & 5 \\ 1 & 3 & 5 & 7 & 2 & 4 & 6 \\ 2 & 4 & 6 & 1 & 3 & 5 & 7 \end{matrix}$$

$h = 3$ $\qquad\qquad\qquad\qquad\qquad\qquad$ $h = 4$
$i = 1,2,3,4,5,6,7$ $\qquad\qquad\qquad\qquad$ $i = 1,2,3,4,5,6,7$
$j = 3,6,2,5,1,4,7$ $\qquad\qquad\qquad\qquad$ $j = 4,1,5,2,6,3,7$

$$A_3 = \begin{matrix} 4 & 7 & 3 & 6 & 2 & 5 & 1 \\ 5 & 1 & 4 & 7 & 3 & 6 & 2 \\ 6 & 2 & 5 & 1 & 4 & 7 & 3 \\ 7 & 3 & 6 & 2 & 5 & 1 & 4 \\ 1 & 4 & 7 & 3 & 6 & 2 & 5 \\ 2 & 5 & 1 & 4 & 7 & 3 & 6 \\ 3 & 6 & 2 & 5 & 1 & 4 & 7 \end{matrix} \qquad A_4 = \begin{matrix} 5 & 2 & 6 & 3 & 7 & 4 & 1 \\ 6 & 3 & 7 & 4 & 1 & 5 & 2 \\ 7 & 4 & 1 & 5 & 2 & 6 & 3 \\ 1 & 5 & 2 & 6 & 3 & 7 & 4 \\ 2 & 6 & 3 & 7 & 4 & 1 & 5 \\ 3 & 7 & 4 & 1 & 5 & 2 & 6 \\ 4 & 1 & 5 & 2 & 6 & 3 & 7 \end{matrix}$$

$h = 5$ $\qquad\qquad\qquad\qquad\qquad\qquad$ $h = 5$
$i = 1,2,3,4,5,6,7$ $\qquad\qquad\qquad\qquad$ $i = 1,2,3,4,5,6,7$
$j = 5,3,1,6,4,2,7$ $\qquad\qquad\qquad\qquad$ $j = 5,3,1,6,4,2,7$

$$A_5 = \begin{matrix} 6 & 4 & 2 & 7 & 5 & 3 & 1 \\ 7 & 5 & 3 & 1 & 6 & 4 & 2 \\ 1 & 6 & 4 & 2 & 7 & 5 & 3 \\ 2 & 7 & 5 & 3 & 1 & 6 & 4 \\ 3 & 1 & 6 & 4 & 2 & 7 & 5 \\ 4 & 2 & 7 & 5 & 3 & 1 & 6 \\ 5 & 3 & 1 & 6 & 4 & 2 & 7 \end{matrix} \qquad A_6 = \begin{matrix} 7 & 6 & 5 & 4 & 3 & 2 & 1 \\ 1 & 7 & 6 & 5 & 4 & 3 & 2 \\ 2 & 1 & 7 & 6 & 5 & 4 & 3 \\ 3 & 2 & 1 & 7 & 6 & 5 & 4 \\ 4 & 3 & 2 & 1 & 7 & 6 & 5 \\ 5 & 4 & 3 & 2 & 1 & 7 & 6 \\ 6 & 5 & 4 & 3 & 2 & 1 & 7 \end{matrix}$$

These are 2, 3, 4, 5, 6-orhtogonal which we can see on superimposing them as follows:

(234567)  (357246)  (473625)  (526374)
(345671)  (461357)  (514736)  (637415)
(456712)  (572461)  (625147)  (741526)
(567123)  (613572)  (736251)  (152637)
(671234)  (724613)  (147362)  (263741)
(712345)  (135724)  (251473)  (374152)
(123456)  (246135)  (362514)  (415263)

(642753)  (765432)  (111111)
(753164)  (176543)  (222222)
(164275)  (217654)  (333333)
(275316)  (321765)  (444444)
(316472)  (432176)  (555555)
(472531)  (543217)  (666666)
(53`642)  (654321)  (777777)

**Conclusive remark 4.1:** Thus when s is prime, by making use of $(1 \pm hj) \bmod s$, $h = 1,2,3,...,(s-1)$, we can get a set of mutually *t*-orthogonal Latin square of order s, where $2 \leq t \leq s-1$ and here $t(N) = s - 1$. The exceptions are for n = 1, 2.

**Type II:** This second type is Latin squares of order s, where (s+1) is a prime number. In this we may have some prime powers and some composite numbers as well.

$$s = 4, 6, 10, 12, 16, 18, 22, 28, 30, 36, 40,\ldots$$

In these cases, we use $a_{ij}^h = (i.j) \bmod (s+1)$, and in the Lain square $A_i$ we obtain, by permuting the rows of $A_i$, such that $\pi A_i = A_{(i+1) \bmod (s+1)}$. We get $A_i s, i = 1,2,...,s$, Latin squares. They form *t*-orthogonal Latin squares where $2 < t \leq s$. We take h = 1 always in this type. These are not of 2-othogonal, but for some of the values of s by some other method we may get 2-orthogonal Latin squares. Let us look at them now.

**Examples 4.2.**

i) for s = 4, as s+1 =5, a prime

$$i = 1,2,3,4 \atop j = 1,2,3,4 \quad \text{and} \quad \pi A_i = A_{(i+1) \bmod 5}.$$

$$A_1 = \begin{matrix} 1234 \\ 2413 \\ 3142 \\ 4321 \end{matrix} \quad A_2 = \begin{matrix} 2413 \\ 3142 \\ 4321 \\ 1234 \end{matrix} \quad A_3 = \begin{matrix} 3142 \\ 4321 \\ 1234 \\ 2413 \end{matrix} \quad A_4 = \begin{matrix} 4321 \\ 1234 \\ 2413 \\ 3142 \end{matrix}$$

These four Latin form 3, 4-othogonal but not 2-othogonal.

$$\begin{matrix} (1234) & (2413) & (3142) & (4321) \\ (2341) & (4132) & (1423) & (3214) \\ (3412) & (1324) & (4231) & (2143) \\ (4123) & (3241) & (2314) & (1432) \end{matrix}$$

In Wallis [25], it was given that, one set of three mutually 2-orthoginal Latin squares of the side 4 is

$$\begin{matrix} 1234 & 1234 & 1234 \\ 2143 & 3412 & 4321 \\ 3412 & 4321 & 2143 \\ 4321 & 2143 & 3412 \end{matrix}$$

Which are being constructed, by using the elements of $GF(p^n)$. They form 2, 3-orthogonal Latin squares as we have

$$\begin{matrix} 111 & 222 & 333 & 444 \\ 234 & 143 & 412 & 321 \\ 342 & 431 & 124 & 213 \\ 423 & 314 & 241 & 132 \end{matrix}$$

**Note 4.1:** In Bose [1], it is given that by using the properties of the Galois Field $GF(p^n)$, it is possible to build up a projective Geometry with $s^2 + s + 1$ points and $s^2 + s + 1$ lines where $s = p^n$, p being a prime integer and s is a positive integer. It has been further shown that the existence of such a geometry is exactly equivalent to the existence of an s-sided completely orthogonalized Latin squares. So he discussed the case in which s = 4, 8, 9, 16, 25, 27,… , which are all prime powers. But here by taking s+1 as prime, we are discussing the cases when s= 4, 16, from the above list, which fall under this Type II.
We have shown above for s=4, in our way we have 3, 4-orhtogonal and in the way of Wallis [25], 2, 3-orthogonal. The other case will be discussed in due course.

**Note 4.2:** Here is another set of 3-orthogonal Latin squares of order 4.

$$\begin{matrix} 1234 & 1234 & 1234 \\ 2413 & 4321 & 3142 \\ 4321 & 3142 & 2413 \\ 3142 & 2413 & 4321 \end{matrix}$$

ii) For s = 6, it is the only Eulerian number that exists. Since n+1 = 7 a prime, we use $(i.j) \bmod 7$, here we have taken h=1 and then by using $\pi A_i = A_{i+1 \bmod 7}$ we get 6 Latin squares, which form 3, 4, 5, 6-orthogonal but not 2-orthogoanl.

$$i = 1,2,3,4,5,6$$
$$j = 1,2,3,4,5,6$$

$$A_1 = \begin{matrix} 123456 \\ 246135 \\ 362514 \\ 415263 \\ 531642 \\ 654321 \end{matrix} \quad A_2 = \begin{matrix} 246135 \\ 362514 \\ 415263 \\ 531642 \\ 654321 \\ 123456 \end{matrix} \quad A_3 = \begin{matrix} 362514 \\ 415263 \\ 531642 \\ 654321 \\ 123456 \\ 246135 \end{matrix}$$

$$A_4 = \begin{matrix} 415263 \\ 531642 \\ 654321 \\ 123456 \\ 246135 \\ 362514 \end{matrix} \quad A_5 = \begin{matrix} 531642 \\ 654321 \\ 123456 \\ 246135 \\ 362514 \\ 415263 \end{matrix} \quad A_6 = \begin{matrix} 654321 \\ 123456 \\ 246135 \\ 362514 \\ 415263 \\ 531642 \end{matrix}$$

On superimposing these we get the array as follows:

(123456) (246135) (362514) (415263) (531642) (654321)
(234561) (461352) (625143) (152634) (316425) (543216)
(345612) (613524) (251436) (526341) (164253) (432165)
(456123) (135246) (514362) (263415) (642531) (321654)
(561234) (352461) (143625) (634152) (425316) (216543)
(612345) (524613) (436251) (341526) (253164) (165432)

But much was said about the Latin squares of order 6. It was well established from the Euler's conjecture [7], and Tarry [23], that there exists no orthogonal mate i.e. 2-orthogonal mate to the Latin square of order 6. At last Hortan [12], gave the construction of two almost orthogonal Latin squares in which by superimposing one on another there are 34 distinct ordered pairs out of 36 pairs. And Heinrich [11] gave the construction for an almost self-orthogonal Latin square of order 6.

So we quote,

**\* From Euler's conjecture 4.1.** No pair of orthogonal Latin squares of order 6 exists, that is $N(6) = 1$.

But now we are in a position to restate that for *t*-orthogonality as follows:

**Proposition 4.1.** There exist t-orthogonal Latin squares of order 6 and $t(6) = 6$, where $t = 3,4,5,6$.

**Proposition 4.2.** In the set of Latin squares of order s where (s+1) is prime the t-orthogonality holds good, where $2 < t \leq s$.

iii) For s=10, by using $(i.j) \mod 11$, and $\pi A_i = A_{i+1 \mod(11)}$ we get 10 Latin squares of order 10, and they form 3,4,5,6,7,8,9,10-orthogonal. Here

$$i = 1,2,3,4,5,6,7,8,9,10$$
$$j = 1,2,3,4,5,6,7,8,9,10$$

And we get the Latin square of order 10 as follows:

| 1 | 2 | 3 | 4 | 5 | 6 | 7 | 8 | 9 | 10 |
|---|---|---|---|---|---|---|---|---|---|
| 2 | 4 | 6 | 8 | 10 | 1 | 3 | 5 | 7 | 9 |
| 3 | 6 | 9 | 1 | 4 | 7 | 10 | 2 | 5 | 8 |
| 4 | 8 | 1 | 5 | 9 | 2 | 6 | 10 | 3 | 7 |
| 5 | 10 | 4 | 9 | 3 | 8 | 2 | 7 | 1 | 6 |
| 6 | 1 | 7 | 2 | 8 | 3 | 9 | 4 | 10 | 5 |
| 7 | 3 | 10 | 6 | 2 | 9 | 5 | 1 | 8 | 4 |
| 8 | 5 | 2 | 10 | 7 | 4 | 1 | 9 | 6 | 3 |
| 9 | 7 | 5 | 3 | 1 | 10 | 8 | 6 | 4 | 2 |
| 10 | 9 | 8 | 7 | 6 | 5 | 4 | 3 | 2 | 1 |

| | | | | |
|---|---|---|---|---|
| (1,2,3,4) | (2,4,6,,8) | (3,6,9,1) | (4,8,1,5) | (5,10,4,9) |
| (2,3,4,5) | (4,6,8,10) | (6,9,1,4) | (8,1,5,9) | (10,4,9,3) |
| (3,4,5,6) | (6,8,10,1) | (9,1,4,7) | (1,5,9,2) | (4,9,3,8) |
| (4,5,6,7) | (8,10,1,3) | (1,4,7,10) | (5,9,2,6) | (9,3,8,2) |
| (5,6,7,8) | (10,1,3,5) | (4,7,10,2) | (9,2,6,10) | (3,8,2,7) |
| (6,7,8,9) | (1,3,5,7) | (7,10,2,5) | (2,6,10,3) | (8,2,7,1) |
| (7,8,9,10) | (3,5,7,9) | (10,2,5,8) | (6,10,3,7) | (2,7,1,6) |
| (8,9,10,1) | (5,7,9,2) | (2,5,8,3) | (10,3,7,4) | (7,1,6,5) |
| (9,10,1,2) | (7,9,2,4) | (5,8,3,6) | (3,7,4,8) | (1,6,5,10) |
| (10,1,2,3) | (9,2,4,6) | (8,3,6,9) | (7,4,8,1) | (6,5,10,4) |
| (6,1,7,2) | (7,3,10,6) | (8,5,2,10) | (9,7,5,3) | (10,9,8,7) |
| (1,7,2,8) | (3,10,6,2) | (5,2,10,7) | (7,5,3,1) | (9,8,7,6) |
| (7,2,8,3) | (10,6,2,9) | (2,10,7,4) | (5,3,1,10) | (8,7,6,5) |
| (2,8,3,9) | (6,2,9,5) | (10,7,4,1) | (3,1,10,8) | (7,6,5,4) |
| (8,3,9,4) | (2,9,5,1) | (7,4,1,9) | (1,10,8,6) | (6,5,4,3) |
| (3,9,4,10) | (9,5,1,8) | (4,1,9,6) | (10,8,6,4) | (5,4,3,2) |
| (9,4,10,5) | (5,1,8,4) | (1,9,6,3) | (8,6,4,2) | (4,3,2,1) |
| (4,10,5,6) | (1,8,4,7) | (9,6,3,8) | (6,4,2,9) | (3,2,1,10) |
| (10,5,6,1) | (8,4,7,3) | (6,3,8,5) | (4,2,9,7) | (2,1,10,9) |
| (5,6,1,7) | (4,7,3,10) | (3,8,5,2) | (2,9,7,5) | (1,10,9,8) |

Furthermore, it can be seen that this gives out 3,4,5,6,7,8,9,10-orthogonal.
Now to consider 2-orthogonality, we adopt the two orthogonal Latin squares constructed by Parker [20, 21].

$$\begin{bmatrix}
10 & 6 & 5 & 4 & 7 & 8 & 9 & 1 & 2 & 3 \\
9 & 1 & 10 & 6 & 5 & 7 & 8 & 2 & 3 & 4 \\
8 & 9 & 2 & 1 & 10 & 6 & 7 & 3 & 4 & 5 \\
7 & 8 & 9 & 3 & 2 & 1 & 10 & 4 & 5 & 6 \\
1 & 7 & 8 & 9 & 4 & 3 & 2 & 5 & 6 & 10 \\
3 & 2 & 7 & 8 & 9 & 5 & 4 & 6 & 10 & 1 \\
5 & 4 & 3 & 7 & 8 & 9 & 6 & 10 & 1 & 2 \\
2 & 3 & 4 & 5 & 6 & 10 & 1 & 7 & 8 & 9 \\
4 & 5 & 6 & 10 & 1 & 2 & 3 & 9 & 7 & 8 \\
6 & 10 & 1 & 2 & 3 & 4 & 5 & 8 & 9 & 7
\end{bmatrix}$$

They form 2-orthogoanl Latin squares, note that by taking any one of the two Latin squares and applying $\pi A_i = A_{i+1 \bmod n}$ does not give $t$-orthogonality. We have already established $t$-orthogonality over the set of Latin squares of order 10.

This construction for a pairs of orthogonal latin squares of order 10. Bose, Shrikhande and Parker [2], and by Menon [16] also gave a pair of orthogonal Latin square of order $n = 3m+1$ for every $m \geq 3$, satisfying $N(m) \geq 2$.

The corresponding theorem (from Parker [21]) states that.

**Theorem 4.1.** [21]**.** Let m be an integer for which there exists a pair of orthogonal Latin squares of order m. Then there exists a pair of orthogonal Latin squares of order $n = 3m+1$.

According for m=1,2,3,4,5,6,7,8,9,10,11,12,13,14 we get n=4,7,10,13,16,19,22,15,28,31,34,37,40,43.
Out of this list 7,13,19,31,37 are of our Type I, and 4,10,1,6,22,28,40, are of the Type II. And 25 is a prime power. Now there remains 34 only, which is to be discussed in Type IV later. Another proposition says that:

**Theorem 4.2:** [Brualdi [4], pp.276]. Let n be a positive integer satisfying n=10 (mod 12). Then there exists a pair of orthogonal Latin squares of order n. It is a corollary of above Theorem 4.1 (cf. [21]). In this the corresponding numbers for n are
$$n=10, 22, 34, 46, 70, 82, 94....$$

Here 10, 22, 46, 70, 82 come under the Type II, 34, 94 are under the type IV to be discussed shortly. Much work was done on the orthogonal Latin squares of order 10, for example Parker [18, 19] has called a Lain square of order n pathological if it cannot form one member of a complete set of mutually orthogonal Latin squares of that order. The Latin square of order 10, is considered to be pathological by him, when 2-orthogonality is under consideration but for *t*-orthogonality when $2 < t \leq 10$. In Franklin [10] triples of almost orthogonal Latin squares of order 10, were constructed, which may be of some use in establishing *t*-orthogonality over them.

e.g. For s-12, since n+1=13, a prime by using $(i.j) \bmod 13$ and $\pi A_i = A_{i+1 \bmod 13}$, we get the Latin square for order 12, as follows. For s=12, $(i.j) \bmod 13$, $\pi A_i = A_{i+1 \bmod 13}$.

$$i = 1,2,3,4,5,6,7,8,9,10,11,12$$
$$j = 1,2,3,4,5,6,7,8,9,10,11,12$$

| 1 | 2 | 3 | 4 | 5 | 6 | 7 | 8 | 9 | 10 | 11 | 12 |
|---|---|---|---|---|---|---|---|---|----|----|----|
| 2 | 4 | 6 | 8 | 10 | 12 | 1 | 3 | 5 | 7 | 9 | 11 |
| 3 | 6 | 9 | 12 | 2 | 5 | 8 | 11 | 1 | 4 | 7 | 10 |
| 4 | 8 | 12 | 3 | 7 | 11 | 2 | 6 | 10 | 1 | 5 | 9 |
| 5 | 10 | 2 | 7 | 12 | 4 | 9 | 1 | 6 | 11 | 3 | 8 |
| 6 | 12 | 5 | 11 | 4 | 10 | 3 | 9 | 2 | 8 | 1 | 7 |
| 7 | 1 | 8 | 2 | 9 | 3 | 10 | 4 | 11 | 5 | 12 | 6 |
| 8 | 3 | 11 | 6 | 1 | 9 | 4 | 12 | 7 | 2 | 10 | 5 |
| 9 | 5 | 1 | 10 | 6 | 2 | 11 | 7 | 3 | 12 | 8 | 4 |
| 10 | 7 | 4 | 1 | 11 | 8 | 5 | 2 | 12 | 9 | 6 | 3 |
| 11 | 9 | 7 | 5 | 3 | 1 | 12 | 10 | 8 | 6 | 4 | 2 |
| 12 | 11 | 10 | 9 | 8 | 7 | 6 | 5 | 4 | 3 | 2 | 1 |

Now we show 3-orthogonality on this by row permutations as defined earlier. 3-orthogonality of Latin squares of order 12 is given by

(1,2,3)    (2,4,6)    (3,6,9)    (4,8,12)   (5,10,2)   (6,12,5)
(2,3,4)    (4,6,8)    (6,9,12)   (8,12,3)   (10,2,7)   (12,5,11)
(3,4,5)    (6,8,10)   (9,12,2)   (12,3,7)   (2,7,12)   (5,11,4)
(4,5,6)    (8,10,12)  (12,2,5)   (3,7,11)   (7,12,4)   (11,4,10)
(5,6,7)    (10,12,1)  (2,5,8)    (7,11,2)   (12,4,9)   (4,10,3)
(6,7,8)    (12,1,3)   (5,8,11)   (11,2,6)   (4,9,1)    (10,3,9)
(7,8,9)    (1,3,5)    (8,11,1)   (2,6,10)   (9,1,6)    (3,9,2)
(8,9,10)   (3,5,7)    (11,1,4)   (6,10,1)   (1,6,11)   (9,2,8)
(9,10,11)  (5,7,9)    (1,4,7)    (10,1,5)   (6,11,3)   (2,8,1)
(10,11,12) (7,9,11)   (4,7,10)   (1,5,9)    (11,3,8)   (8,1,7)
(11,12,1)  (9,11,2)   (7,10,3)   (5,9,4)    (3,8,5)    (1,7,6)
(12,1,2)   (11,2,4)   (10,3,6)   (9,4,8)    (8,5,10)   (7,6,12)

(7,1,8)    (8,3,11)   (9,5,1)    (10,7,4)   (11,9,7)   (12,11,10)
(1,8,2)    (3,11,6)   (5,1,10)   (7,4,1)    (9,7,5)    (11,10,9)
(8,2,9)    (11,6,1)   (1,10,6)   (4,1,11)   (7,5,3)    (10,9,8)
(2,9,3)    (6,1,9)    (10,6,2)   (1,11,8)   (5,3,1)    (9,8,7)
(9,3,10)   (1,9,4)    (6,2,11)   (11,8,5)   (3,1,12)   (8,7,6)
(3,10,4)   (9,4,12)   (2,11,7)   (8,5,2)    (1,12,10)  (7,6,5)
(10,4,11)  (4,12,7)   (11,7,3)   (5,2,12)   (12,10,8)  (6,5,4)
(4,11,5)   (12,7,2)   (7,3,12)   (2,12,9)   (10,8,6)   (5,4,3)
(11,5,12)  (7,2,10)   (3,12,8)   (12,9,6)   (8,6,4)    (4,3,2)
(5,12,6)   (2,10,5)   (12,8,4)   (9,6,3)    (6,4,2)    (3,2,1)
(12,6,7)   (10,5,8)   (8,4,9)    (6,3,10)   (4,2,11)   (2,1,12)
(6,7,1)    (5,8,3)    (4,9,5)    (3,10,7)   (2,11,9)   (1,12,11)

Further we can easily establish 4, 5, 6, 7, 8, 9, 1 0, 11, 12-orthogonality over this Latin square of order 12. But it was states in Macneish [14, pp.222] that "The simplest case would be to prove that the Euler square of index 12, 3 is impossible". By the above example we are constrained to contradict it now. For this s =12, Lain square Johnson, Dulmage and Mendelshon [13], gave he construction of 5 MOLS of order 12, and Wallis [26] prove that $N(12) \geq 5$ for 2-orthogonality. By taking the 5 Lain squares from [13], we can establish 2, 3, 4, 5 –orthogonality over them and thus we have established $t$-orthogonality where $2 < t \leq 12$, here.

**Conclusive remark 4.2:** Similarly we can consider for s=16, 18, 22… and in each of these cases we may find some other way of dealing 2-orthogonality as above.

**Type III:** Now we consider prime powers

$$s = 4, 8, 16, 32, 64; \quad 9, 27, 81; \quad 25, 125; \quad 49, 121\ldots$$

Among these under Type III, we have already considered 4, 16. Now we consider the remaining one by one. In fact we construct Latin squares for any number, as in Type I, but we should to take into consideration $(h, s) = 1$, for getting Latin squares. Still orthogonality is the question, even I these Latin squares.

**Examples 4.3.**

i) For s=8, the Latin squares can constructed with the formula $(i + hj) \mod n$, where $(h, s) = 1$. So here h= 1, 3, 5, 7.

$$h = 1$$
$$i = 1,2,3,4,5,6,7,8$$
$$j = 1,2,3,4,5,6,7,8$$

$$A_1 = \begin{matrix} 23456781 \\ 34567812 \\ 45678123 \\ 56781234 \\ 67812345 \\ 78123456 \\ 81234567 \\ 12345678 \end{matrix}$$

$$h = 3$$
$$i = 1,2,3,4,5,6,7,8$$
$$j = 3,6,1,4,7,2,5,8$$

$$A_2 = \begin{matrix} 47258361 \\ 58361472 \\ 61472583 \\ 72583614 \\ 83614725 \\ 14725836 \\ 25836147 \\ 36147258 \end{matrix}$$

$$h = 5$$
$$i = 1,2,3,4,5,6,7,8$$
$$j = 5,2,7,4,1,6,3,8$$

$$A_3 = \begin{matrix} 63852741 \\ 74163852 \\ 85274163 \\ 16385274 \\ 27416385 \\ 38527416 \\ 41638527 \\ 52741638 \end{matrix}$$

$$h = 7$$
$$i = 1,2,3,4,5,6,7,8$$
$$j = 7,6,5,4,3,2,1,8$$

$$A_4 = \begin{matrix} 87654321 \\ 18765432 \\ 21876543 \\ 32187654 \\ 43218765 \\ 54321876 \\ 65432187 \\ 76543218 \end{matrix}$$

In these there exists no *t*-orthogonality $2 \leq t \leq 8$. But from Bose [1], if we take the construction of the Latin squares of order 8, and by applying $\pi A_i = A_{(i+1) \mod 8}$, we can get 4, 5, 6, 7, 8-orthogonal Latin squares but not 2, 3-orthogonal. Let us see that construction for s = 8.

|   |   |   |   |   |   |   |   |
|---|---|---|---|---|---|---|---|
| 8 | 1 | 2 | 3 | 4 | 5 | 6 | 7 |
| 1 | 8 | 6 | 4 | 3 | 7 | 2 | 5 |
| 2 | 6 | 8 | 7 | 5 | 4 | 1 | 3 |
| 3 | 4 | 7 | 8 | 1 | 6 | 5 | 2 |
| 4 | 3 | 5 | 1 | 8 | 2 | 7 | 6 |
| 5 | 7 | 4 | 6 | 2 | 8 | 3 | 1 |
| 6 | 2 | 1 | 5 | 7 | 3 | 8 | 4 |
| 7 | 5 | 3 | 2 | 6 | 1 | 4 | 8 |

Construction for t-orthogonality is as follows:

| 8123 | 1864 | 2687 | 3478 | 4351 | 5746 | 6215 | 7532 |
|------|------|------|------|------|------|------|------|
| 1234 | 8643 | 6875 | 4781 | 3518 | 7462 | 2157 | 5326 |
| 2345 | 6437 | 8754 | 7816 | 5182 | 4628 | 1573 | 3261 |
| 3456 | 4372 | 7541 | 8165 | 1827 | 6283 | 5738 | 2614 |
| 4567 | 3725 | 5413 | 1652 | 8276 | 2831 | 7384 | 6148 |
| 5678 | 7251 | 4132 | 6523 | 2764 | 8315 | 3846 | 1487 |
| 6781 | 2518 | 1326 | 5234 | 7643 | 3157 | 8462 | 4875 |
| 7812 | 5186 | 3268 | 2347 | 6435 | 1574 | 4621 | 8753 |

This is 4-orthogonal and hence *t*-orthogonal for *t* = 4, 5, 6, 7, 8.
As in Bose [1], keeping the first row as it is and applying on the other rows $\pi R_i = R_{(i+1) \mod 8}$ formula, we get another six Latin squares and these 7 Latin squares from MOLS of order 8. Thus we have

| (88) | (11) | (22) | (33) | (44) | (55) | (66) | (77) |
|------|------|------|------|------|------|------|------|
| (12) | (86) | (68) | (47) | (35) | (74) | (21) | (53) |
| (23) | (64) | (87) | (78) | (51) | (46) | (15) | (32) |
| (34) | (43) | (75) | (81) | (18) | (62) | (57) | (26) |
| (45) | (37) | (54) | (16) | (82) | (28) | (73) | (61) |
| (56) | (72) | (41) | (65) | (27) | (83) | (38) | (14) |
| (67) | (25) | (13) | (52) | (76) | (31) | (84) | (48) |
| (71) | (58) | (36) | (24) | (63) | (17) | (42) | (85) |

2-orthogonal and hence, we can built up 3, 4, 5, 6, 7-orthogonal in the similar way. And thus on Latin squares of order 8, we have t-orthogonality where $2 \leq t \leq 8$. Franklin [9] had constructed the Latin square of order 8.

e.g. For s = 9, a prime power with $(i + hj) \mod 9$, we can get Latin squares of order 9, for h = 1,2,4,5,7,8. But the orthogonality is not possible.

From Bose [1], if we take the construction of Latin squares of order 9, we have

| 9 | 1 | 2 | 3 | 4 | 5 | 6 | 7 | 8 |
| 1 | 5 | 8 | 4 | 6 | 9 | 3 | 2 | 7 |
| 2 | 8 | 6 | 1 | 5 | 7 | 9 | 4 | 3 |
| 3 | 4 | 1 | 7 | 2 | 6 | 8 | 9 | 5 |
| 4 | 6 | 5 | 2 | 8 | 3 | 7 | 1 | 9 |
| 5 | 9 | 7 | 6 | 3 | 1 | 4 | 8 | 2 |
| 6 | 3 | 9 | 8 | 7 | 4 | 2 | 5 | 1 |
| 7 | 2 | 4 | 9 | 1 | 8 | 5 | 3 | 6 |
| 8 | 7 | 3 | 5 | 9 | 2 | 1 | 6 | 4 |

By applying on this $\pi R_i = R_{i+1 \mod 9}$, we get 9 Latin squares superimposing $t$ of them we obtain we can establish $t$-orthogonality where $t = 3,4,5,6,7,8,9$. Let us see 3-orthogonality with this process.

| 912 | 158 | 286 | 341 | 465 | 597 | 639 | 724 | 873 |
| 128 | 584 | 861 | 417 | 652 | 976 | 398 | 249 | 735 |
| 284 | 846 | 615 | 172 | 528 | 763 | 987 | 491 | 359 |
| 845 | 469 | 157 | 726 | 283 | 631 | 874 | 918 | 592 |
| 456 | 693 | 579 | 268 | 837 | 314 | 742 | 185 | 921 |
| 567 | 932 | 794 | 689 | 371 | 148 | 425 | 853 | 216 |
| 678 | 327 | 943 | 895 | 719 | 482 | 251 | 536 | 164 |
| 789 | 271 | 432 | 953 | 194 | 825 | 516 | 367 | 648 |
| 891 | 715 | 328 | 534 | 946 | 259 | 163 | 672 | 487 |

This is 3-orthogonal and by continuing the process we can further establish t-orthogonality where t= 4, 5, 6, 7, 8, 9

And for 2-orthogonality, by keeping the first row as it is and applying on the remaining 8 row $\pi R_i = R_{i+1 \mod 9}$, we get 7 other Latin squares, which are 2-othogonal. Lets us see.

| | | | | | | | | |
|---|---|---|---|---|---|---|---|---|
| (99) | (11) | (22) | (33) | (44) | (55) | (66) | (77) | (88) |
| (12) | (58) | (86) | (41) | (65) | (97) | (39) | (24) | (73) |
| (28) | (84) | (61) | (17) | (52) | (76) | (98) | (49) | (35) |
| (84) | (46) | (15) | (72) | (28) | (63) | (87) | (91) | (59) |
| (45) | (69) | (57) | (26) | (83) | (31) | (74) | (18) | (92) |
| (56) | (93) | (79) | (68) | (37) | (14) | (42) | (85) | (21) |
| (67) | (32) | (94) | (89) | (71) | (48) | (25) | (53) | (16) |
| (78) | (27) | (43) | (95) | (19) | (82) | (51) | (36) | (64) |
| (81) | (75) | (38) | (54) | (96) | (29) | (13) | (62) | (47) |

We can further establish *t*-orthogonality on this where t=3, 4, 5, 6, 7, 8.

**Conclusive remark 4.3.** From Bose [1], as he has already dealt with the cases of prime powers i.e. for s=25, 27…, we can establish t-orthogonality similarly, on them as above cases of 8, 9 and 16 of Type II.

**Type IV:** Now we consider the remaining numbers say

$$s=14, 15, 20, 21, 24, 26, 33, 34, 35, 38, 39.$$

In all these cases by following the method of Type I, where $(h, s) = 1$, we can construct Latin squares. But, the orthogonality is to be established and that seems to be difficult.

For this type of composite numbers, Macneish [14] and Mann [15], have given the methods of constructions of Latin squares.

For the Latin square of order 14, given by Wallis [pp.197, 26], which is an idempotent Latin square, i.e. the principle diagonal is as (1, 2, … , 14), and by applying $\pi R_i = R_{(i+1)} \mod 14$, we get *t*-orthogonality where *t*=5,6,7,8,…,14. This is self orthogonal (orthogonal to its transpose). The cyclic generation of 2-orthogonal of these self orthogonal Latin squares was discussed by Franklin [8, 9]. From [26] we have

| | | | | | | | | | | | | | |
|---|---|---|---|---|---|---|---|---|---|---|---|---|---|
| 1 | 9 | 4 | 13 | 10 | 3 | 6 | 1 | 17 | 12 | 2 | 5 | 14 | 8 |
| 14 | 2 | 10 | 5 | 1 | 11 | 4 | 7 | 12 | 8 | 13 | 3 | 6 | 9 |
| 7 | 14 | 3 | 11 | 6 | 2 | 12 | 5 | 8 | 13 | 9 | 1 | 4 | 10 |
| 5 | 8 | 14 | 4 | 12 | 7 | 3 | 13 | 6 | 9 | 1 | 10 | 2 | 11 |
| 3 | 6 | 9 | 14 | 5 | 13 | 8 | 4 | 1 | 7 | 10 | 2 | 11 | 12 |
| 12 | 4 | 7 | 10 | 14 | 6 | 1 | 9 | 5 | 2 | 8 | 11 | 3 | 13 |
| 4 | 13 | 5 | 8 | 11 | 14 | 7 | 2 | 10 | 6 | 3 | 9 | 12 | 1 |
| 13 | 5 | 1 | 6 | 9 | 12 | 14 | 8 | 3 | 11 | 7 | 4 | 10 | 2 |
| 11 | 1 | 6 | 2 | 7 | 10 | 13 | 14 | 9 | 4 | 12 | 8 | 5 | 3 |
| 6 | 12 | 2 | 7 | 3 | 8 | 11 | 1 | 14 | 10 | 5 | 13 | 9 | 4 |
| 10 | 7 | 13 | 3 | 8 | 4 | 9 | 12 | 2 | 14 | 11 | 6 | 1 | 5 |
| 2 | 11 | 8 | 1 | 4 | 9 | 5 | 10 | 13 | 3 | 14 | 12 | 7 | 6 |
| 8 | 3 | 12 | 9 | 2 | 5 | 10 | 6 | 11 | 1 | 4 | 14 | 13 | 7 |
| 9 | 10 | 11 | 12 | 13 | 1 | 2 | 3 | 4 | 5 | 6 | 7 | 8 | 14 |

By applying $\pi R_i = R_{(i+1) \bmod 14}$ we get 13 other Latin squares and by superimposing them, first we have

| | | | | | | |
|---|---|---|---|---|---|---|
| 1,14,7 | 9,2,14,8 | 4,10,3 | 13,5,11 | 10,1,6 | 3,11,2 | 6,4,12 |
| 14,7,5,3,12 | 2,14,8 | 10,3,14 | 5,11,4 | 1,6,12 | 11,2,7 | 4,12,3 |
| 7,5,3 | 14,8,6 | 3,14,9 | 11,4,14 | 6,12,5 | 2,7,13 | 12,3,8 |
| 5,3,12 | 8,6,4 | 14,9,7 | 4,14,10 | 12,5,14 | 7,13,6 | 3,8,1 |
| 3,12,4 | 6,14,13 | 9,7,5 | 14,10,8 | 5,14,11 | 13,6,14 | 8,1,7 |
| 12,4,,13 | 4,13,5 | 7,5,1 | 10,8,6 | 14,11,9 | 6,14,12 | 1,7,14 |
| 4,13,11 | 13,5,1 | 5,1,6 | 8,6,2 | 11,9,7 | 14,12,10 | 7,14,13 |
| 13,11,6 | 5,1,12 | 1,6,2 | 6,2,7 | 9,7,3 | 12,10,8 | 14,13,11 |
| 11,6,10 | 1,12,7 | 6,2,13 | 2,7,3 | 7,3,8 | 10,8,4 | 13,11,9 |
| 6,10,2 | 12,7,11 | 2,13,8 | 7,3,1 | 3,8,4 | 8,4,9 | 11,9,5 |
| 10,2,8 | 7,11,3 | 13,8,12 | 3,1,9 | 8,4,2 | 4,9,5 | 9,5,10 |
| 2,8,9 | 11,3,10 | 8,12,11 | 1,9,12 | 4,2,13 | 9,5,1 | 5,10,2 |
| 8,9,1 | 3,10,9 | 12,11,4 | 9,12,13 | 2,13,10 | 5,1,3 | 10,2,6 |
| 9,1,14 | 10,9,2 | 11,4,10 | 12,13,5 | 13,10,1 | 1,3,11 | 2,6,4 |
| 11,7,5 | 7,12,8 | 12,8,13 | 2,13,9 | 5,3,1 | 14,6,4 | 8,9,10 |
| 7,5,13 | 12,8,6 | 8,13,9 | 13,9,1 | 3,1,10 | 6,4,2 | 9,10,11 |
| 5,13,4 | 8,6,1 | 13,9,7 | 9,1,10 | 1,10,2 | 4,2,11 | 10,11,12 |
| 13,4,9 | 6,1,5 | 9,7,2 | 1,10,8 | 10,2,11 | 2,11,3 | 11,12,13 |
| 4,9,2 | 1,5,10 | 7,2,6 | 10,8,3 | 2,11,9 | 11,3,12 | 12,13,1 |
| 9,2,8 | 5,10,3 | 2,6,11 | 8,3,7 | 11,9,4 | 3,12,10 | 13,1,2 |
| 2,8,14 | 10,3,9 | 6,11,4 | 3,7,12 | 9,4,8 | 12,10,5 | 1,2,3 |
| 8,14,1 | 3,9,14 | 11,4,10 | 7,12,5 | 4,8,13 | 10,5,9 | 2,3,4 |
| 14,1,12 | 9,14,2 | 4,10,14 | 12,5,11 | 8,13,6 | 5,9,1 | 3,4,5 |
| 1,12,10 | 14,2,13 | 10,14,3 | 5,11,14 | 13,6,12 | 9,1,7 | 4,5,6 |
| 12,10,6 | 2,13,11 | 14,3,1 | 11,14,4 | 6,12,14 | 1,7,13 | 5,6,7 |
| 10,6,3 | 13,11,4 | 3,1,5 | 14,4,6 | 12,14,7 | 7,13,8 | 6,7,14 |
| 6,3,11 | 11,4,7 | 1,5,12 | 4,6,10 | 14,7,5,3,1 | 13,8,14 | 7,14,8 |
| 3,11,7 | 4,7,12 | 5,12,8 | 6,2,13 | 7,5,3 | 8,14,6 | 14,8,9 |

**Note 4.3:** Only at the stage of 5-orthogonal all tuples become distinct for example the tuple (14, 7, 5, 3, 1) and tuple (14, 7, 5, 3, 12), which are distinct 5-tuples. So it is 5-orthogonal. Further we can show 6, 7, 8,…, 14-orthogonal.

By retaining the first row and in the remaining rows by applying $\pi R_i = R_{(i+1) \mod 14}$ we get 4-orthogonal. From the construction of Todorov [24] who has constructed 3 mutually orthogonal Latin squares of order 14, we can establish 2,3-orthogonality also.

| | | | | | | |
|---|---|---|---|---|---|---|
| 1,1,1 | 9,9,9 | 4,4,4 | 13,13,13 | 10,10,10 | 3,3,3 | 6,6,6 |
| 14,7,5 | 2,14,8 | 10,3,14 | 5,11,4 | 1,6,12 | 11,2,7 | 4,12,3 |
| 7,5,12 | 14,8,6 | 3,14,9 | 11,4,14 | 6,12,5 | 2,7,13 | 12,3,8 |
| 5,3,12 | 8,6,4 | 14,9,7 | 4,14,10 | 12,5,14 | 7,13,6 | 3,8,1 |
| 3,12,4 | 6,4,13 | 9,7,5 | 14,10,8 | 5,14,11 | 13,6,14 | 8,1,7 |
| 12,4,13 | 4,13,5 | 7,5,1 | 10,8,6 | 14,11,9 | 6,14,12 | 1,7,14 |
| 4,13,11 | 13,5,1 | 5,1,6 | 8,6,2 | 11,9,7 | 14,12,10 | 7,14,13 |
| 13,11,6 | 5,1,12 | 1,6,2 | 6,2,7 | 9,7,3 | 12,10,8 | 14,13,11 |
| 11,6,10 | 1,12,7 | 6,2,13 | 2,7,3 | 7,3,8 | 10,8,4 | 13,11,9 |
| 6,10,2 | 12,7,11 | 2,13,8 | 7,3,1 | 3,8,4 | 8,4,9 | 11,9,5 |
| 10,2,8 | 12,7,11 | 2,13,8 | 7,3,1 | 3,8,4 | 8,4,9 | 11,9,5 |
| 10,2,8 | 2,11,3 | 13,8,12 | 3,1,9 | 8,4,2 | 4,9,5 | 9,5,10 |
| 2,8,9 | 11,3,10 | 8,12,11 | 1,9,12 | 4,2,13 | 9,5,1 | 5,10,2 |
| 5,9,1 | 3,10,9 | 12,11,4 | 9,12,13 | 2,13,,10 | 5,1,3 | 10,2,6 |
| 9,1,4 | 10,9,2 | 11,4,10 | 12,13,5 | 13,10,1 | 1,3,11 | 2,6,4 |
| 11,11,11   | 7,7,7     | 12,12,12 | 2,2,2   | 5,5,5    | 14,14,14 | 8,8,8    |
| 7,5,13     | 12,8,6    | 8,13,9   | 13,9,1  | 3,1,10   | 6,4,2    | 9,10,11  |
| 5,13,4     | 8,6,1     | 13,9,7   | 9,1,10  | 1,10,2   | 4,2,11   | 10,11,12 |
| 13,4,9     | 6,1,5     | 9,7,2    | 1,10,8  | 10,2,11  | 2,11,3   | 11,12,13 |
| 4,9,2      | 1,5,10    | 7,2,6    | 10,8,3  | 2,11,9   | 11,3,12  | 12,13,1  |
| 9,2,8      | 5,10,3    | 2,6,11   | 8,3,7   | 11,9,4   | 3,12,10  | 13,1,2   |
| 2,8,14     | 10,3,9    | 6,11,4   | 3,7,12  | 9,4,8    | 12,10,5  | 1,2,3    |
| 8,14,1     | 3,9,14    | 11,14,10 | 7,12,5  | 4,8,13   | 10,5,9   | 2,3,4    |
| 14,1,12    | 9,14,2    | 4,10,14  | 12,5,11 | 8,13,6   | 5,9,1    | 3,4,5    |
| 1,12,10    | 14,2,13   | 10,14,3  | 5,11,14 | 13,6,12  | 9,1,7    | 4,5,6    |
| 12,10,6    | 2,13,11   | 14,3,1   | 11,14,4 | 6,12,14  | 1,7,13   | 5,6,7    |
| 10,6,3     | 13,11,4   | 3,1,5    | 14,4,6  | 12,11,7  | 7,13,8   | 6,7,14   |
| 6,3,11     | 11,4,7    | 1,5,12   | 4,6,2   | 14,7,5   | 13,8,14  | 7,14,8   |
| 3,11,7     | 4,7,12    | 5,12,8   | 6,2,13  | 7,5,3    | 8,14,1   | 14,8,9   |

For n = 15. In Wallis [pp.82, 26], stated that $N(15) \geq 4$. And the first rows of these squares are

| 1 | 15 | 2  | 14 | 3 | 13 | 4  | 12 | 5  | 11 | 6  | 10 | 7  | 9  | 8  |
| 1 | 14 | 3  | 11 | 6 | 9  | 8  | 7  | 10 | 4  | 13 | 12 | 5  | 15 | 2  |
| 1 | 10 | 7  | 13 | 4 | 2  | 15 | 6  | 11 | 9  | 8  | 3  | 14 | 12 | 5  |
| 1 | 6  | 11 | 10 | 7 | 15 | 2  | 5  | 12 | 14 | 3  | 9  | 8  | 4  | 13 |

The later rows of each square are obtained by adding 1, 2,…, 14 to every element modulo (15). These help us in establishing the *t*-orthogonality.

For n = 20. Macneish [14] gave the construction of orthogonal Latin square of order 20, where $N(20) = 3$; which easily establish 2 and 3- orthogonality. By using $\pi A_i = A_{i+1}$ on any of these, we can further establish *t*-orthogonality.

Furthermore, from Raghavarao [pp.44, 45; 22] we have the tabulation of MOLS up to s = 100. From that we have

$$N(21) = 2, \quad N(24) = 2, \quad N(26) = 1, \quad N(33) = 2,$$
$$N(34) = 1, \quad N(35) = 4, \quad N(38) = 1, \ldots\ldots\ldots$$

And taking these constructions we can discuss t-orthogonality in the similar way.

Wallis [25] states that

**Theorem 4.3.** There exist there mutually orthogonal Latin squares of every order except 2, 3, 6 and possibly 10 and 14.

We have established t-orthogonality for 3,6,10 and 14 here.

Wallis [25], further states that

**Theorem 4.4.** There exist three mutually orthogonal Latin squares of order 18, 22, 26, 30, 34, 38 and 42. Among these for 18, 22, 30 and 42, as they are of Type II, we have established t-orthogonality where $t > 2$. And for 26, 34, and 38 as there are 3 mutually orthogonal Latin squares as they are 2-orhtogonal we can establish 3-orthogonality and possibly *t*-orthogonality by taking any one of the Latin squares and applying $\pi R_i = R_{(i+1) \bmod n}$.

**Note 4.4:** To stress once again in a *t*-tuple element may repeat, by any *t*-tuple should not be repeated in the whole of the array for having *t*-orthogonality.
Bose, Shrikhande [3], established, while proving the falsity of Euler's conjecture, the non existence of two orthogonal Latin squares of order $4t + 2$. In this series we have 6,10,14,18,22,26,30,34 and for 6,10,18,22,31 they come under Type II, for 14 we have already established *t*-orthogonality and for 26 Wallis [25] established that there exist 3 MOLS of order 26.

In correspondence to these orthogonalities, i.e. 2-orthogonality and *t*-orthogonality the status of Latin squares of each order are discussed in a table here under.

Lastly, what are the necessary and sufficient conditions for the existence of *t*-orthogonality over a given set of orthogonal Latin squares is open for further research investigation.
The use of this *t*-orthogonality is in the construction of certain SBIB and SPIB designs and some t-partite graphs can be seen in a sequel to this paper.

|  | 2-orthogonality | | | | t-orthogonality | |
| --- | --- | --- | --- | --- | --- | --- |
| Order of the Latin squares | Number of MOLS existing | Method of construction (Reference) | Number of M t-O LS existing | | t= | Method of construction (Reference) |
| 1 | ∞ | Trivial | ∞ | | 1 | Trivial Type I |
| 2 | 1 | Trivial | 1 | | t=1 | Trivial Type I |
| 3 | 2 | Prime, Trivial, Bose [1] | 2 | Prime | t=2 | Trivial Type I |
| 4 | 3 | Prime, Trivial, Bose [1] | 4 | 4+1=5 | t=3,4 | Type II |
|  | 2 | Bose, Parker and Shrikhande [2] | 3 | | t=2,3 | Bose [1], Wallis [p. 184, 25] |
|  | 2 | Menon [16] | 3 | | t=3 only | (Constructed in this paper) |
| 5 | 4 | Prime Bose [1] | 4 | Prime | t=2,3,4 | Type I |
| 6 | 1 | Horton [12], 2, almost orthogonal Latin squares Tarry [23], Heinrich [11] almost self orthogonal | 6 | 6+1=7 | t=3,4,5,6 | Type II |
| 7 | 6 | Bose [1], prime | 6 | Prime | t=2,3,4,5, 6 | Type I |
|  | 2 | Bose, Parker and Shrikhande [2] Menon [16] | | | | |
| 8 | 7 | Bose [1], Franklin [9] | 7 | Prime power | t=2,3,4,5, 6,7 t=4,5,6,7 | Bose [1] Type III |
| 9 | 8 | Prime power, Bose [1] | 9 | Prime power | t=3…9 | Type III |
|  | | | 7 | | t=2…7 | Bose [1] |
| 10 | 2 | Franklin [10] triples of almost orthogonal Parker [18,19], Brualdi [4, p.276], Bose, Parker and Shrikhande [2] Menon [16] Parker, [18,19] pathological | 10 | 10+1=11 | t=3…10 | Type II |
| 11 | 10 | Prime Bose [1] | 11 | Prime | t=2…10 | Type I |

|  | 2-orthogonality | | | | t-orthogonality | |
|---|---|---|---|---|---|---|
| Order of the Latin squares | Number of MOLS existing | Method of construction (Reference) | Number of M t-OLS existing | t= | | Method of construction (Reference) |
| 12 | 5 | Johnson, Dulmage and Mendelshon [13] | 12 | 12+1=13 | t=3,…,12 t=2,3,4,5,6 | Type II Johnson, Dulmage and Mendelshon [13] |
|  | 5 | Wallis [26] |  |  |  | Wallis [26] |
| 13 | 12 | Prime, Bose [1] Bose, Parker, and shrikhande [2], Menon [16] | 12 | Prime | t=2…12 | Type I |
| 14 | 3 | Wallis [26] self orthogonal Franklin [8] Todorov [24] | 14 |  | t=5…14 t=2…3 | Trivial Type IV Todorov [21] Wallis [25] |
| 15 | 2 | Macniesh-Mann Theorem [14-15] | 4 |  | t=2,3,4 | Type IV |
|  | 4 | Wallis [25] |  |  |  | Wallis [25] |
| 16 | 2 | Bose, Parker and Shrikhande [2], Menon [16] | 16 | 18+1=19 | t=3…16 | Type II |
| 17 | 16 | Prime, Bose [1] | 16 | Prime | t=2…16 | Type I |
| 18 | 3 | Wallis [25] | 18 | Prime | t=2,3 t=3…18 | Wallis [25] Type II |
| 19 | 18 | Prime, Bose [1] Bose, Parker and Shrikhande [2], Menon [16] | 18 | Prime | t=2…18 | Type I |
| 20 | 3 | Macniesh-Mann Theorem [14-15] | 3 |  | t=2,3 | Macniesh-Mann Theorem [14-15] Type IV |

| | 2-orthogonality | | | | t-orthogonality | |
|---|---|---|---|---|---|---|
| Order of the Latin squares | Number of MOLS existing | Method of construction (Reference) | Number of M t-OLS existing | | $t=$ | Method of construction (Reference) |
| 21 | 2 | Raghavarao [22] | 2 | | $t=2$ | Raghavarao [22] |
| 22 | 2 | Bose, Parker, and shrikhande [2], Menon [16] Brualdi [4 p-276] | 22 | Prime | $t=3…22$ | Type II |
| | 3 | Wallis [25] | 3 | | $t=2,3$ | Wallis [25] |
| 23 | 22 | Prime, Bose [1] | 22 | Prime | $t=2…22$ | Type I |
| 24 | 2 | Raghavarao [22] | 2 | | $t=2$ | Raghavarao [22] Type IV |
| 25 | 2 | Prime power Bose, Parker and Shrikhande [2], Menon [16] | 2 | | $t=2$ | Type III |
| 26 | 3 | Wallis [25] | 3 | Prime | $t=2,3$ | Wallis [25] Type IV |
| 27 | 26 | Prime power Bose [1] | 26 | Prime | $t=2…6$ | Bose [1] Type II |
| 28 | 2 | Bose, Parker and Shrikhande [2], Menon [16] | 28 | 28+1=29 | $t=3…28$ | Type II |

| | 2-orthogonality | | | | $t$-orthogonality | |
|---|---|---|---|---|---|---|
| Order of the Latin squares | Number of MOLS existing | Method of construction (Reference) | Number of M $t$-OLS existing | | $t=$ | Method of construction (Reference) |
| 29 | 28 | Prime, Bose [1] | 28 | Prime | $t=2\ldots28$ | Type I |
| 30 | 3 | Wallis [25] | 30 | 30+1=31 | $t=3\ldots20$ $t=2,3$ | Type II Wallis [25] |
| 31 | 30 2 | Prime, Bose [1] Bose, Parker and Shrikhande [2], Menon [16] | 30 | Prime | $t=2\ldots30$ | Type I |
| 32 | 31 | Prime power, Bose [1] | 31 | Prime power | $t=3\ldots31$ | Type III |
| 33 | 2 | Raghavarao [22] | 2 | | $t=2$ | Raghavarao [22] Type I |
| 34 | 2 | Macneish-Mann Theorem [14-15] | 4 | | | |
| 35 | 4 | Macneish-Mann Theorem [14-15] | 4 | | $t=3\ldots35$ | Type II Macneish-Mann Theorem [14-15] |
| 36 | 3 | Macneish-Mann Theorem [14-15] | 36 | 36+1=37 | $t=3\ldots35$ $t=2,3$ | Type II Macneish-Mann Theorem [14-15] |

| | 2-orthogonality | | | | t-orthogonality | |
|---|---|---|---|---|---|---|
| Order of the Latin squares | Number of MOLS existing | Method of construction (Reference) | Number of M t-O LS existing | | t= | Method of construction (Reference) |
| 37 | 36 | Prime, Bose [1] | 36 | Prime | $t=2…36$ | Prime, Type I |
|  | 2 | Bose, Parker and Shrikhande [2], Menon [16] | | | | |
| 38 | 3 | Wallis [25] | 3 | | $t=2,3$ | Wallis [25], Type IV |
| 39 | 2 | | 2 | | $t=2$ | Raghavarao [22] |
| 40 | 4 | Macneish-Mann Theorem [14-15] | 40+1=41 | $t=2,3$ | $t=3…40$ | Type II Macneish-Mann |
| 41 | 40 | Prime, Bose[1] | 40 | Prime | $t=2…40$ | Type I |
| 42 | 3 | Wallis [25] | 3 | 42+1=43 | $t=3…42$ $t=2,3$ | Type II Wallis [25] |
| 43 | 42 | Prime, Bose [1] Bose, Parker and Shrikhande [2], Menon [16] | 42 | Prime | $t=t…42$ | Type I |
| 44 | 2 3 | Menon [16] Macneish-Mann [14-15] | 3 | | $t=2,3$ | Macneish-Mann Theorem [14-15]Type IV |
| 45 | 4 | Macneish-Mann Theorem [14-15] | 4 | | $t=2,3,4$ | Macneish-Mann Theorem [14-15]Type IV |

**Acknowledgements:** One of the authors Mohan is thankful to Prof. M.G. K. Menon, who is a fountainhead of inspiration to him and to the Third World Academy of Sciences, Trieste, Italy and Prof. Bill Chen, Center for Combinatorics, Nankai University, Tianjin, PR China, for giving him an opportunity to work in the center in China for three months, twice during 2005 and 2006. Mohan also wish to thank Prof. P.Fan, of Jiangtong University, PR China, where the work has been finalized this paper. His thanks are also due to Sir CRR College authorities namely Administrative Officer K.Srimanarayana, Principal R.Surya Rao, and Secretary Gutta Subbarao for their kind support in his research quest. He is also thankful to Prof. Moon Ho Lee for extending invitation to visit Chonbuk National University, South Korea.

This work was partially supported by the MIC (Ministry of Information and Communication), under the ITFSIP (IT Foreign Specialist Inviting Program ) supervised by IITA, under ITRC supervised by IITA, and International Cooperation Research Program of the Ministry of Science & Technology, Chonbuk National University, Korea and partially by the Third World Academy of Sciences, Italy, and Jingtong University, PR China. Hence all the concerned authorities are gratefully acknowledged